\documentclass[useAMS,usenatbib,usegraphicx]{mn2e}

\newcommand{\kev}{keV}
\newcommand{\ergcms}{erg~cm$^{-2}$~s$^{-1}$}
\newcommand{\fe}{Fe~K$\alpha$}
\newcommand{\fxfd}{$F_x/F_{\mathrm{disc}}$}
\newcommand{\etal}{et al.}
\newcommand{\mcg}{MCG--6-30-15}

\title[Fe~K$\alpha$ response from accretion discs]
  {The response of the Fe~K$\mathbf{\alpha}$ line to changes in the
  X-ray illumination of accretion discs}
\author[D.\ R.\ Ballantyne \&  R.\ R.\ Ross]
  {D.~R.~Ballantyne$^1$\thanks{drb@ast.cam.ac.uk} and  R.~R.~Ross$^2$\\
  $^1$Institute of Astronomy, Madingley Road, Cambridge CB3 0HA \\
  $^2$Physics Department, College of the Holy Cross, Worcester, MA 01610, USA}

\pagerange{\pageref{firstpage}--\pageref{lastpage}}
\pubyear{2001}

\usepackage{times}

\begin{document}

\label{firstpage}

\maketitle

\begin{abstract}
X-ray reflection spectra from photoionized accretion discs in active
galaxies are presented for a wide range of illumination
conditions. The energy, equivalent width (EW) and flux of the \fe\
line are shown to depend strongly on the ratio of illuminating flux to
disc flux, \fxfd, the photon index of the irradiating power-law,
$\Gamma$, and the incidence angle of the radiation, $i$. When \fxfd$
\leq 2$ a neutral \fe\ line is prominent for all but the largest
values of $\Gamma$. At higher illuminating fluxes a He-like \fe\ line
at 6.7~\kev\ dominates the line complex. With a high-energy cutoff of
100~\kev, the thermal ionization instability seems to suppress the
ionized \fe\ line when $\Gamma \leq 1.6$. The \fe\ line flux
correlates with \fxfd, but the dependence weakens as iron becomes
fully ionized. The EW is roughly constant when \fxfd\ is low and a
neutral line dominates, but then declines as the line progresses
through higher ionization stages. There is a strong positive
correlation between the \fe\ EW and $\Gamma$ when the line energy is
at 6.7~\kev, and a slight negative one when it is at 6.4~\kev. This is
a potential observational diagnostic of the ionization state of the
disc. Observations of the broad \fe\ line which take into account any
narrow component would be able to test these predictions. Ionized \fe\
lines at 6.7~\kev\ are predicted to be common in a simple magnetic
flare geometry. A model which includes multiple ionization gradients
on the disc is postulated to reconcile the results with observations.
\end{abstract}

\begin{keywords}
accretion, accretion discs -- line: profiles -- galaxies: active -- 
X-rays: general
\end{keywords}

\section{Introduction}
\label{sect:intro}
The discovery of the iron K$\alpha$ line and Compton reflection in the
X-ray spectra of accreting black holes was an important step in the
understanding of the central engine \citep*{mu93}. These features were
predicted to result from the reprocessing of X-rays by an optically
thick and cold medium \citep*{lw88,gr88,geo91,ma91}, quite likely the
accretion disc itself. This was spectacularly confirmed by the
\textit{ASCA} observation of a broad \fe\ line in the X-ray spectrum
of the bright Seyfert~1 galaxy \mcg\ \citep{tan95}. The profile of
this line was well fit by a model of line emission from
relativistically moving material within 10 Schwarszchild radii of a
supermassive black hole \citep*{fab89}. Alternative explanations for
such a broad line suffer from physical inconsistencies and/or
fine-tuning of model parameters \citep*{fab95,reyw00}. Suddenly a
potentially powerful probe of accretion and black hole physics was
observationally accessible \citep*{fab00}. Many other active galactic
nuclei (AGN) were subsequently observed by \textit{ASCA} in search of
a broad \fe\ line, and, although most of the detections were of far
less quality than the one of \mcg, the mean line profile of a sample
of Seyferts seemed to be broadened in a similar manner \citep*{n97a,yaq02}.

Since AGN themselves have quite variable X-ray continua, it was
anticipated that the \fe\ line should also change over an
observation. The largest dataset has come from long observations of
\mcg\ where the line is seen to sometimes drastically change energy
\citep*{iwa96,iwa99}, and its flux varies independently of the
continuum \citep{ve01}. This last property, also seen in other objects
\citep*{wan99,ngm99,chi00,wan01,wea01}, is contrary to the predictions
of the simple reflection scenario, which predicts the line variations
should track the continuum. The fact that this is not observed
suggests a more complicated and dynamic pattern of illumination on the
disc \citep{rey00} such as in the model of irradiation by magnetic
flares in a patchy corona \citep*{gal79,haa91,haa94,mf01}. These
flares may be rotating \citep{rus00}, outflowing \citep*{bel99,mbp01}
and/or temporally connected \citep*{pf99,mf01}, and so they will
clearly have an impact on the observed \fe\ line due to the changing
pattern of radiation on the surface of the disc.

Many models of X-ray reflection from AGN-like accretion discs have
been published \citep*[e.g.][]{ros93,zyk94,mz95,ros99}, but they have
not progressed to the same dynamic level as the models of accretion
disc coronae. The best current models compute the reflection spectrum
from a photoionized layer on the surface of an accretion disc in
hydrostatic equilibrium \citep*{nkk00,brf01,rdc01}. If discs are anything
like what standard theory predicts, then these calculations should be
able to make specific predictions about reflection spectra
\citep[e.g.][]{nay00,nk01}.

Here, we compute such spectra from an AGN accretion disc over a wide
range of illumination conditions, and examine the behaviour of the
energy, equivalent width (EW) and flux of the \fe\ line. Our approach
differs from that of \citet{nk01} or \citet{zyk01} by concentrating on
the observationally accessible changes to the \fe\ line.  The
parameters are chosen so that conditions appropriate for both flaring
and quiescent regions of the disc are explored. Comparison of the
model predictions with time-averaged data (to avoid any
non-equilibrium effects) may allow constraints to be placed on how
discs are irradiated.

The paper is structured as follows. Section~\ref{sect:comp} describes
the model, the assumptions, and the range of parameters that are
treated. The results of the calculations and a discussion of the
evolution of the \fe\ line profile, flux and EW are given in
Section~\ref{sect:res}. A discussion of the results is given in
Section~\ref{sect:discuss} before Section~\ref{sect:concl} summarizes
the primary conclusions.

\section{Computations}
\label{sect:comp}
We employ the code of \citet{ros93}, recently extended by \citet{brf01}, to
calculate the X-ray reflection spectrum from the ionized layer on the
surface of an AGN accretion disc. The details of the calculations are
described in the above two papers (see also \citet{ro78} and
\citealt*{rwm78}), and so only a brief description is given here.

The model calculates the thermal, ionization and density structure of
the top five\footnote{When \fxfd$=270$ a total Thomson depth of 10 was
used to extend the computational domain beyond the layer that is
highly ionized by the illuminating radiation.} Thomson depths of an
accretion disc that is irradiated by an external source of X-rays from
above, and a soft blackbody from the disc below. The calculation is
one-dimensional, with the height of the bottom of the atmosphere given
by the disc equations of \citet*{mfr00} in the gas-pressure dominated
regime and assuming $\alpha=0.1$. The flux of soft radiation entering
the layer from below is assumed to result from viscous heating deep in
the disc and is calculated using the standard equation from
\citet{sha73}. The impinging hard X-rays are in the form of a
power-law with photon index $\Gamma$ (so that photon flux $\propto
E^{-\Gamma}$), which strikes the surface with net flux $F_x$ and at an
angle of $i$ degrees to the normal. The incident X-rays extend from
1~eV and terminate with a sharp cutoff at 100~\kev.

The incident radiation is transferred analytically using a one-stream
approximation \citep[e.g.,][]{ryl79}. Diffuse radiation within the
layer (i.e., incident X-rays that have undergone Compton scattering,
other emission from the gas itself, and the soft photons working their
way outward from the disc below) is treated via the
Fokker-Planck/diffusion approximation \citep{rwm78}, while the \fe\
line is treated in detail via escape and destruction probabilities
assuming a Lorentz line profile (see \citealt{ro78},
\citealt{ros93}). Hydrogen and helium are assumed to be fully ionized
everywhere within the layer, and the following ionization stages of
the most abundant metals are treated: C~{\sc v--vii}, O~{\sc v--ix},
Mg~{\sc ix--xiii}, Si~{\sc xi--xv}, and Fe~{\sc xvi--xxvii}. The
elemental abundances are given by \citet{mcm83}.

The major differences between this code and the one presented by
\citet{nkk00} are in the method of radiative transfer (we use
Fokker-Planck/diffusion, as opposed to a variable Eddington factor
scheme) and in the level of detail that atomic physics is treated
(\citet{nkk00} employ \textsc{xstar} for the photoionization
calculations and so includes many more transitions than our
code). However, comparisons between the two programs yield only minor
qualitative differences in the output spectra, particularly around the
\fe\ line \citep{peq02}.

To isolate the \fe\ variations due to a changing illumination pattern,
the accretion disc parameters were set to representative values. A
typical black hole mass of 10$^8$~M$_{\odot}$ \citep{geb00} was used
in the calculations, as was an accretion rate of 0.001 of the
Eddington rate, which seems typical of some X-ray bright Seyfert 1
galaxies \citep{rf01}. A radius of 7 Schwarszchild radii was selected
for where the reflection takes place, but we do not take into account
relativistic blurring. With these parameters, the soft flux entering
the atmosphere from below is $F_{\mathrm{disc}}= 1.2 \times
10^{13}$~\ergcms, and it is assumed that there is no energy
dissipation in the corona. Other choices for these system parameters would
result in different \fe\ EWs and fluxes, but the resulting trends with
the illumination parameters should not be affected (Sect.~\ref{sub:consist}).

Models were calculated for $1.5 \leq \Gamma \leq 2.1$, which is the
range most often observed for Seyfert~1 galaxies \citep{n97a}, and
with the incident flux covering \fxfd$\approx$ 0.5 -- 270. To
test for geometrical effects (which may be important for magnetic
flares where the size of the hard X-ray emitting region is much
smaller than the disc), for each value of $\Gamma$ and $F_x$ models
were calculated for X-ray incidence angles $20 \leq i \leq 80$~degrees
in steps of 20~degrees. Finally, we neglect the effects of inclination
along the line of sight, but since this does not depend on the disc or
illumination model, earlier results \citep*[e.g.,][]{mfr96} will still
apply.

\section{Results}
\label{sect:res}

\subsection{\fe\ line profiles}
\label{sub:lines}
The properties of the \fe\ line depend greatly on the ionization state
of the gas (e.g., \citealt*{mfr93,mfr96}; \citealt{ros99}). When the
gas is only weakly ionized or neutral (so that Fe~{\sc i--xvi} are the
dominant ionization states), a strong line appears at 6.4~\kev\ and
the associated absorption edge is close to 7.1~\kev. As the gas becomes more
ionized, intermediate Fe states such as Fe~{\sc xvii--xxiii} are
dominant. However, the K$\alpha$ lines from these ions are
preferentially suppressed due to Auger destruction \citep*{rfb96}. It
is not until the gas becomes so ionized that He-like iron (Fe~{\sc
xxv}) is the dominant species that the observed K$\alpha$ line shifts
in energy. This ion has a large fluorescent yield resulting in a
strong line and absorption edge at 6.7~\kev\ and $\sim 8.5$~\kev,
respectively\footnote{\citet{oel01} showed that under certain
conditions the dielectronic satellite lines of Fe~{\sc xxv} could move
the energy centroid of the line closer to 6.6~\kev.}. The line from
hydrogenic Fe at 6.97~\kev\ is also present in these situations, but
is weaker than the helium-like line because of resonant
scattering. Finally, it is possible for Fe to be fully stripped and
neither a line nor an edge will be imprinted on the reflection
spectrum.

In variable density models it is not possible to describe the spectrum
by assigning an ionization parameter to it. However, the general
evolution of the \fe\ line that was described above is seen in these
new calculations, but the addition of a realistic density structure
can create significant departures from this trend.
\begin{figure}
\includegraphics[width=0.50\textwidth]{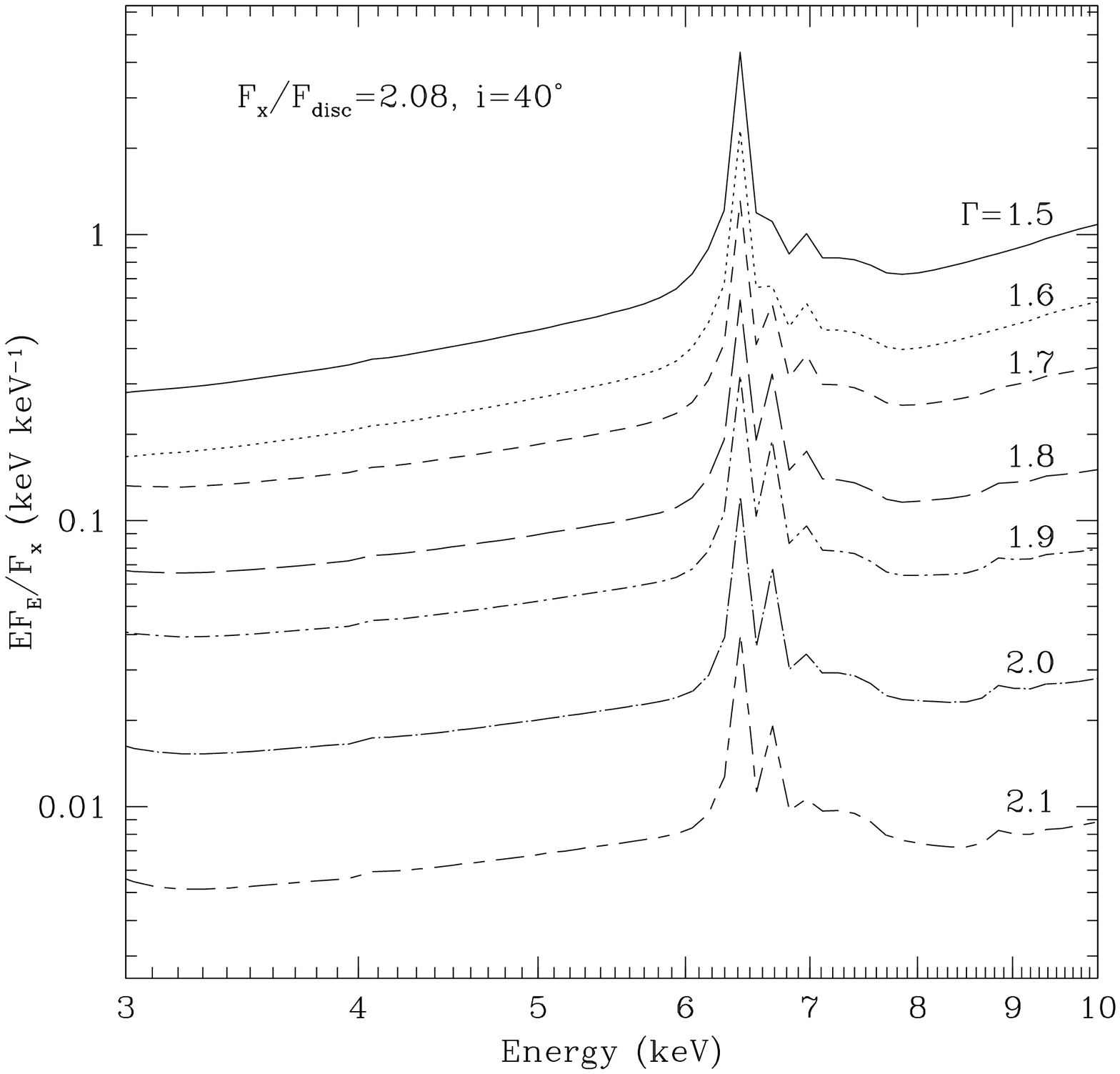}
\includegraphics[width=0.50\textwidth]{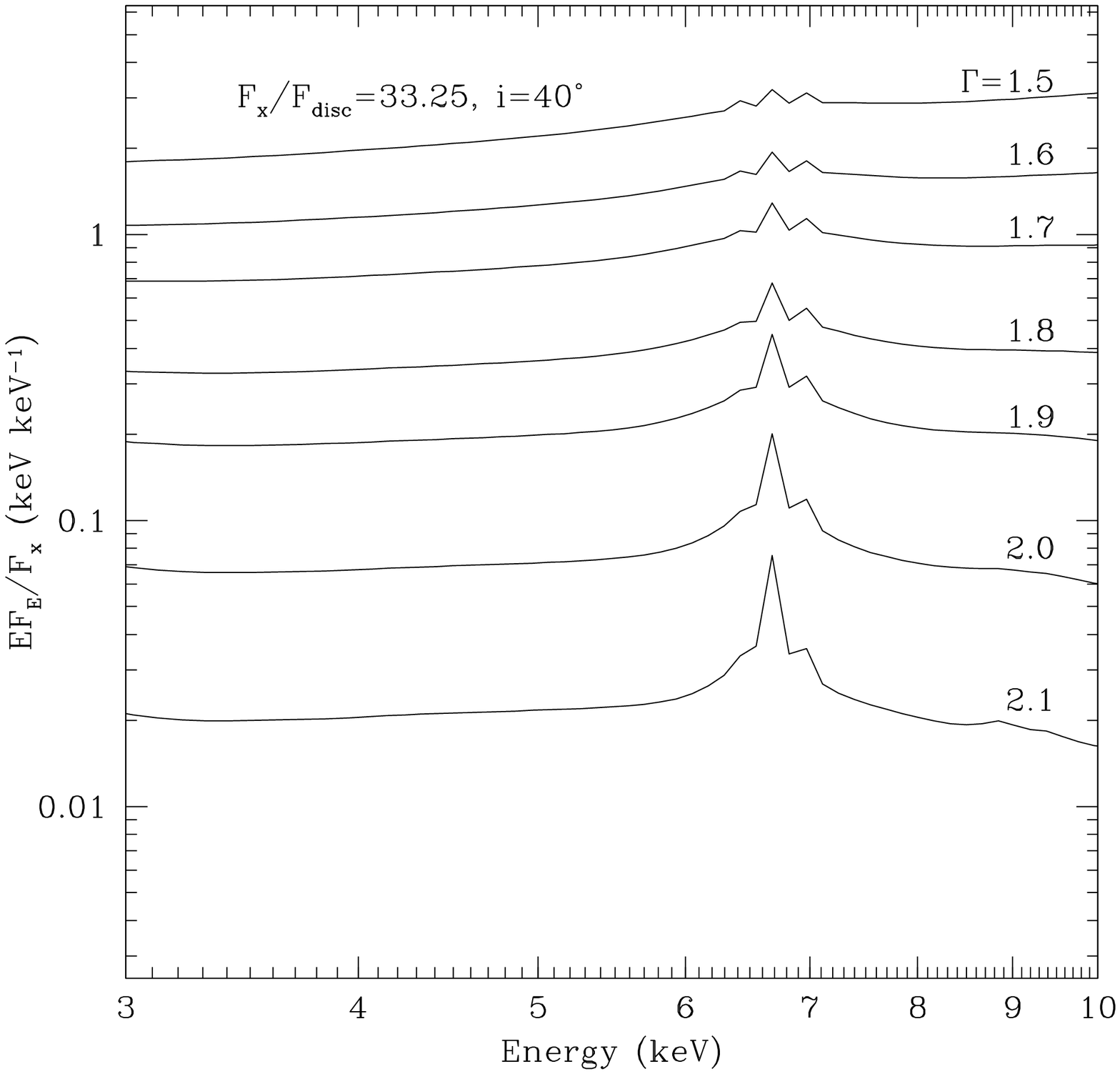}
\caption{\textit{Top:} The variation in the \fe\ line profile due to
  changes in the photon index of the illuminating radiation. $\Gamma$
  increases from 1.5 at the top of the plot to 2.1 at the bottom, and
  the spectra have been vertically offset for clarity. \textit{Bottom:}
  The same plot as in the other panel, but at a higher value of
  \fxfd. Here, the \fe\ line becomes more ionized at lower $\Gamma$,
which is opposite to the previous case.}
\label{fig:gamma-line}
\end{figure}
Figure~\ref{fig:gamma-line} shows how the \fe\ region of the
reflection spectrum can vary for two different values of \fxfd\ when
the photon index of the illuminating spectrum changes. At the lower
value of \fxfd, the \fe\ emission lines become more ionized (i.e., has
a significant He-like component) with larger values of $\Gamma$ than
with smaller values. This is possibly counter-intuitive because the
harder irradiating spectra have more Fe-ionizing power than the softer
ones. However, when \fxfd\ is increased (the bottom panel of
Fig.~\ref{fig:gamma-line}) the situation is reversed, and the \fe\
line becomes highly ionized and weak at the lowest values of $\Gamma$.

The explanation for this behaviour is illustrated 
\begin{figure}
\includegraphics[width=0.50\textwidth]{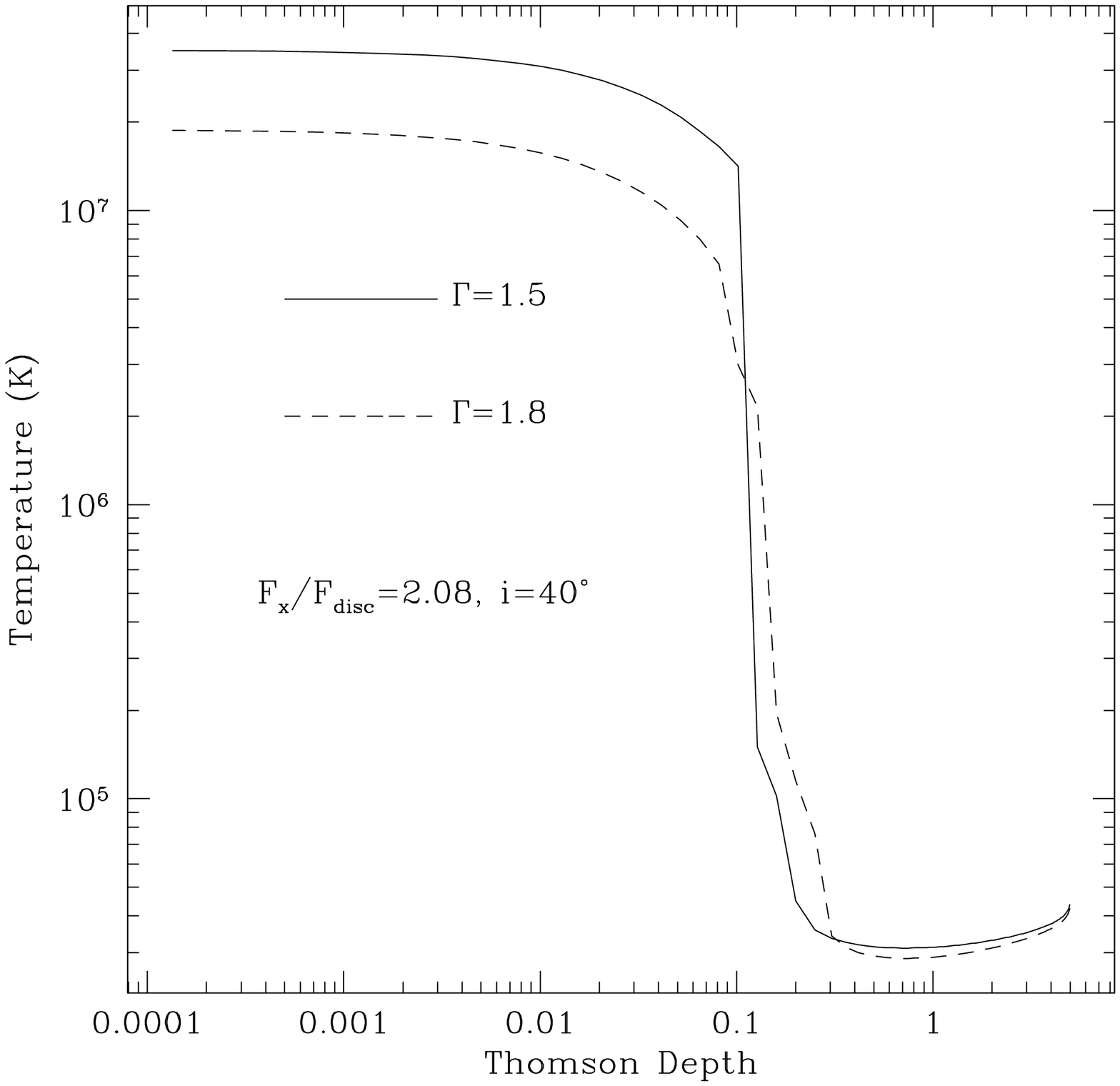}
\includegraphics[width=0.50\textwidth]{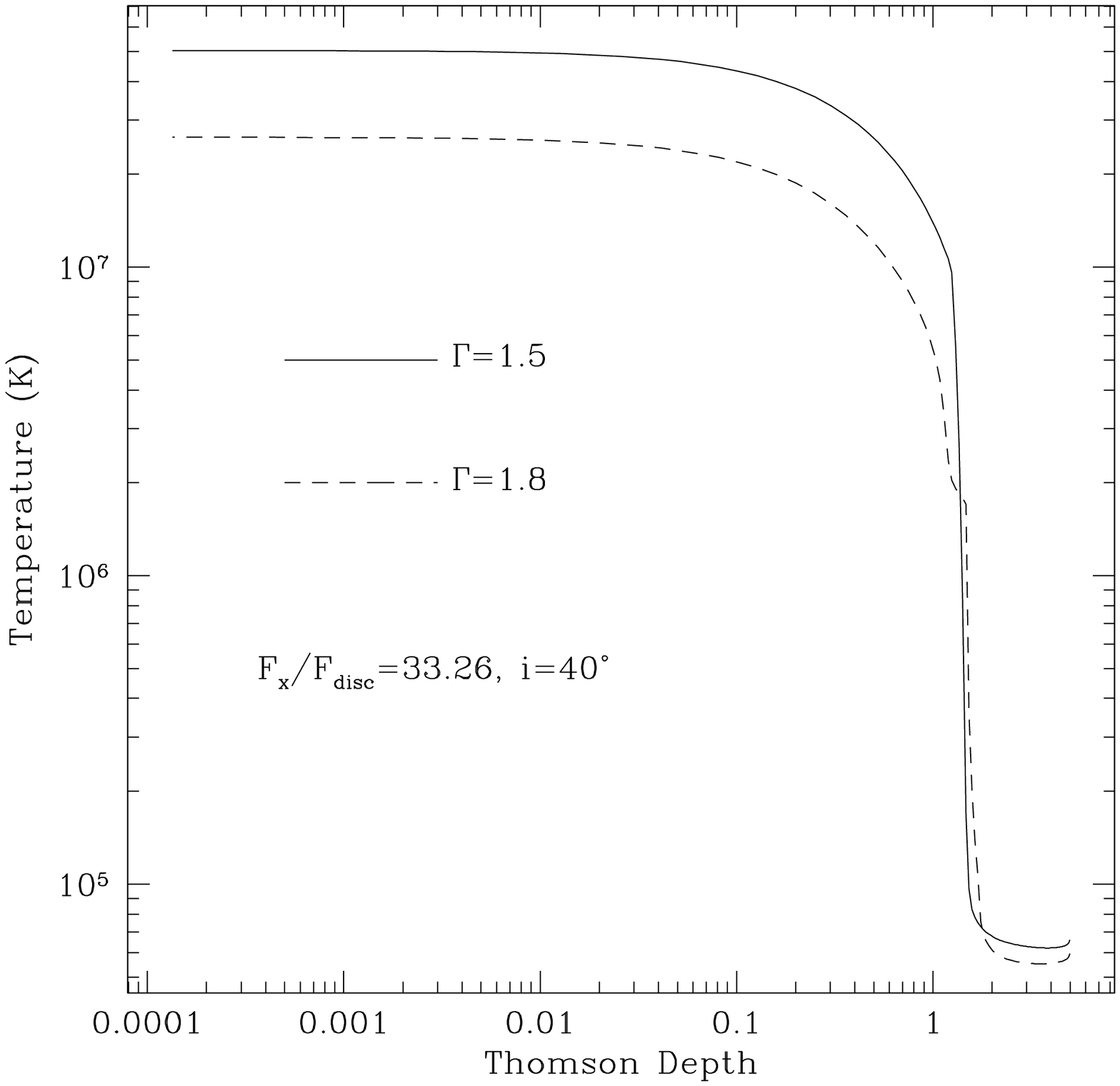}
\caption{Temperature vs.\ Thomson Depth plots for the $\Gamma=1.5$ and
1.8 models that were shown in Fig.~\ref{fig:gamma-line}. When
$\Gamma=1.5$ the temperature falls almost non-continuously from a hot,
ionized phase to a cold, recombined phase which is attributed to the
thermal ionization instability. As \fxfd\ is increased, the
transition occurs at a non-negligible Thomson depth, and less ionizing
photons can reach neutral gas without scattering. The $\Gamma=1.8$
illuminating spectrum cannot heat the gas to as high a temperature,
and so the instability does not occur.}
\label{fig:temp-tau}
\end{figure}
in Figure~\ref{fig:temp-tau}. The plots show that when $\Gamma=1.5$
the temperature of the photoionized gas falls abruptly from a very hot
($\sim$ Compton temperature) fully ionized phase to a cold, recombined
phase. \citet{nkk00} showed that such behaviour is due to the
well-known thermal ionization instability of gases in pressure balance
(e.g., \citealt*{kmt81}). There is a negligible fraction of gas with
intermediate ionization states, so, when \fxfd\ is relatively small,
only a neutral \fe\ line is emitted with any real
strength\footnote{Thermal conduction between the two temperature zones
may be important under some conditions, allowing enough gas at
intermediate temperatures to imprint spectral features on the
reflection spectrum \citep*{roz99,li01}.}. At higher values of \fxfd\
the incident radiation ionizes further into the atmosphere and the
transition occurs at a non-negligible Thomson depth. Few K-shell
ionizing photons can traverse this layer to the neutral material below
without being scattered outwards or absorbed by a higher ionized
species of iron. Therefore, at higher values of \fxfd, the line
weakens and disappears at the lowest values of $\Gamma$. As the
photon-index is increased, the maximum temperature of the gas
decreases (\citealt*{gfm83} and Sect.~\ref{sub:instability}) and the
instability cannot operate. Other ionized species of Fe can exist near
the surface and a He-like line appears in the spectrum. From
Fig.~\ref{fig:gamma-line} we see that one effect of this instability is to suppress ionized \fe\ lines when $\Gamma \leq 1.6$ (but see
Sect.~\ref{sub:instability}).

A typical example of how the \fe\ line evolves with increasing \fxfd\
is shown
\begin{figure}
\includegraphics[width=0.50\textwidth]{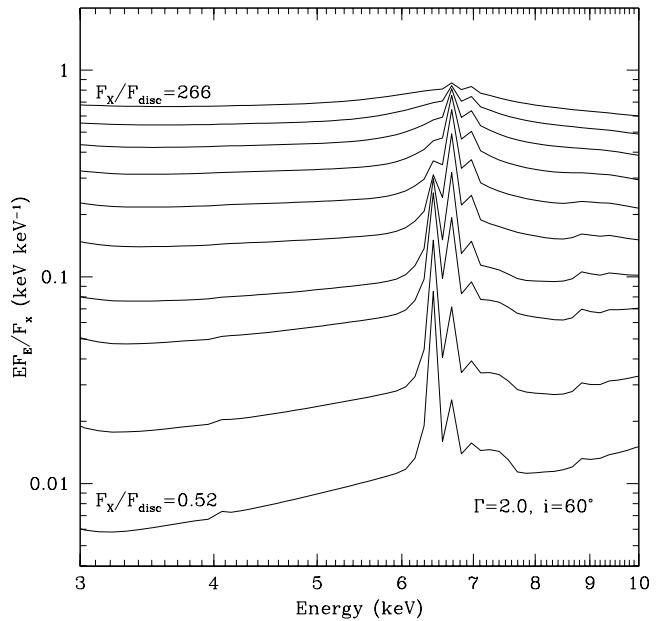}
\caption{The evolution of the \fe\ line with \fxfd\ for $\Gamma=2.0$
  and an incidence angle of $60$~degrees. \fxfd$=0.52$ for the lower
  spectrum and increases by factors of two to 270 for the upper
  spectrum. The spectra have been offset vertically for clarity. The
  energy evolution of the line is reminiscent of the one predicted by the
  constant density reflection models.}
\label{fig:flux-line}
\end{figure}
in Figure~\ref{fig:flux-line}. When \fxfd=0.5 or 1.0 then the surface
of the accretion disc is not highly ionized and a neutral line at
6.4~\kev\ dominates the line profile. As \fxfd\ is increased the
He-like line strengthens and then weakens as the atmosphere is further
ionized. Also, the line profile gradually broadens due to Compton
scattering. This trend is similar to the predictions of constant
density reflection models \citep{ros99}.

Finally, we will briefly discuss the impact of the incidence angle of
the incoming radiation.
\begin{figure}
\includegraphics[width=0.50\textwidth]{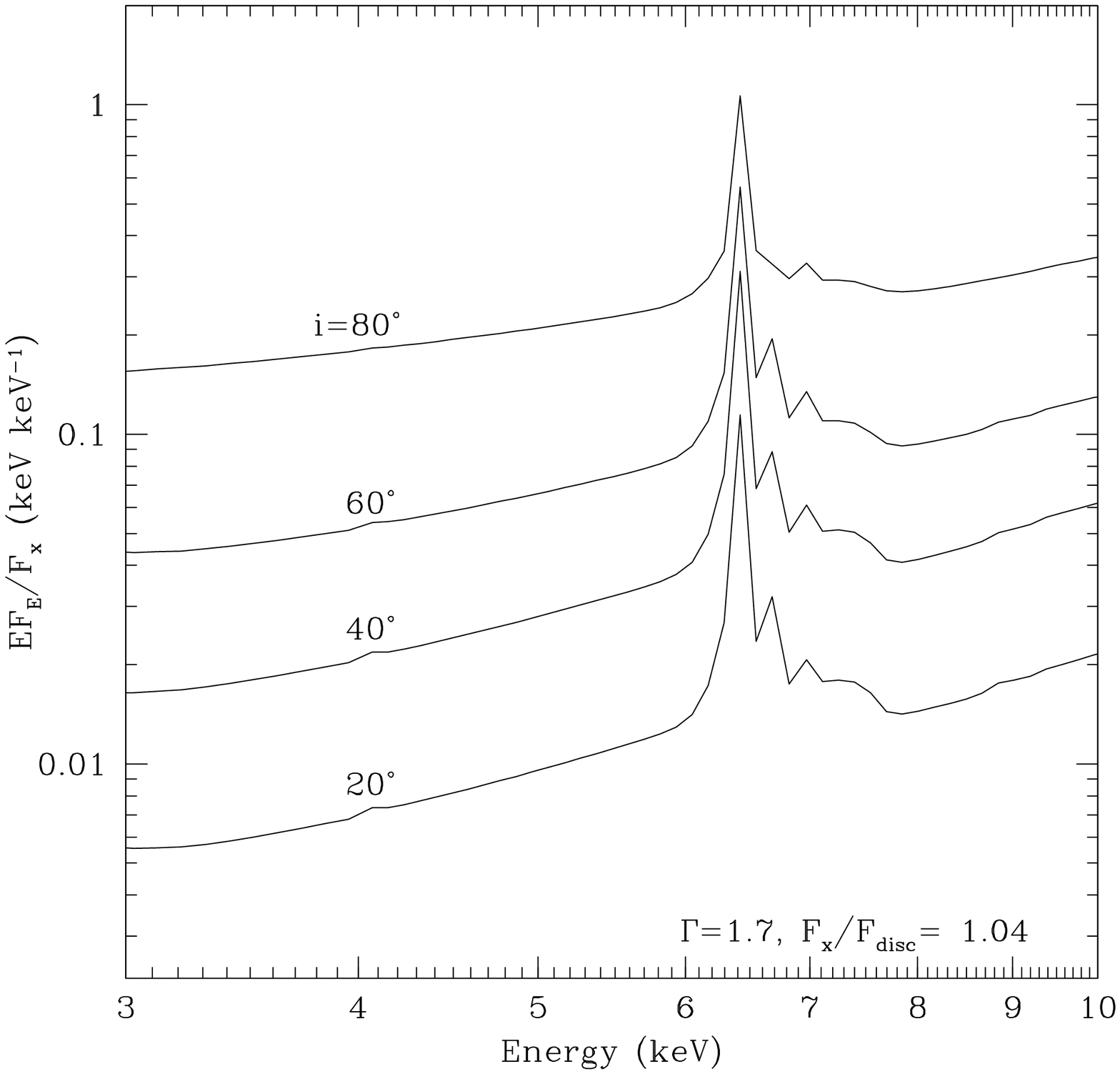}
\includegraphics[width=0.50\textwidth]{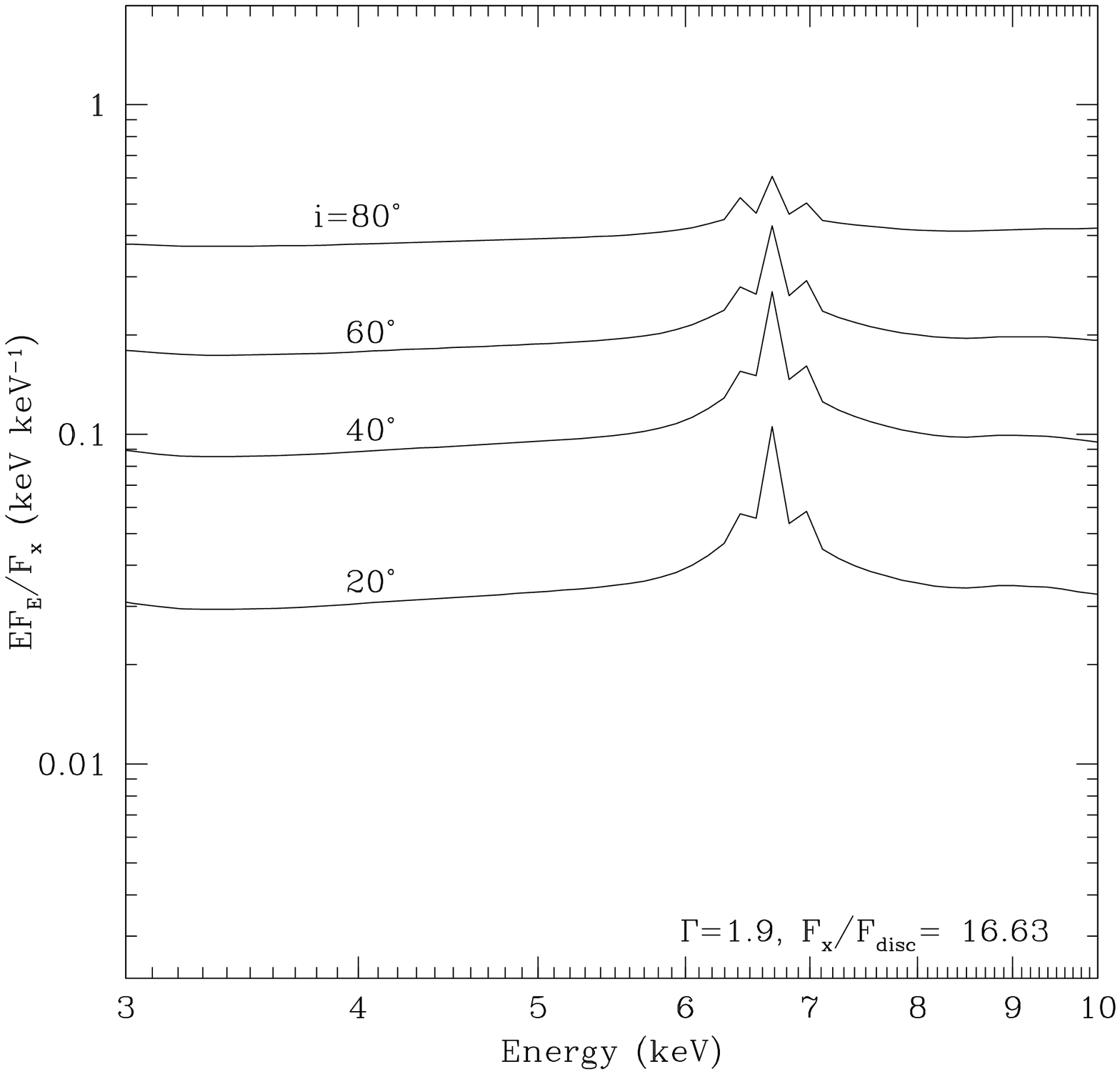}
\caption{\textit{Top:} The variation in the \fe\ line profile due to
  changes in the incidence angle of the illuminating radiation. $i$
  increases from 20~degrees at the bottom of the plot to 80~degrees at
  the top, and the spectra have been vertically offset for
  clarity. \textit{Bottom:} The same plot as in the other panel, but
  at higher values of \fxfd\ and $\Gamma$. The incidence angle has the
  strongest effect when \fxfd\ is low, as the temperature of the gas
  increases with $i$.}
\label{fig:angle-line}
\end{figure}
When the angle of the beam is increased, the radiation encounters any
specific line-of-sight Thomson depth at a smaller physical depth into the
atmosphere. Therefore, the energy is dumped into a smaller physical region,
which increases the temperature of the gas \citep[see
also][]{nkk00}. When \fxfd\ is relatively small this can enhance the
thermal instability, as is shown in the top panel of
Figure~\ref{fig:angle-line}. In that case, the He-like component in
the line profile disappears when $i=80$~degrees. However, when \fxfd\
is greater (the bottom panel of Fig.~\ref{fig:angle-line}) the line
photons are emitted at a significant Thomson depth, and the hot
outer layers result in a weaker overall line profile.

\subsection{Equivalent width and line flux}
\label{sub:ew-lf}
The two most important observational diagnostics of the strength of
the \fe\ line are its equivalent width (EW) and the flux in the
line. Therefore, it is important to investigate how they vary as the
illumination of the disc is changed. To facilitate comparison with
observations, the illuminating power-law was added to the reflection
spectrum before the measurements were made. This can be thought of as
a reflection fraction of unity, where the reflection fraction measures
the strength of the reflected component in an observed spectrum. In
the disc-corona model of an AGN, reflection fractions close to one are
expected.

\begin{figure}
\includegraphics[width=0.50\textwidth]{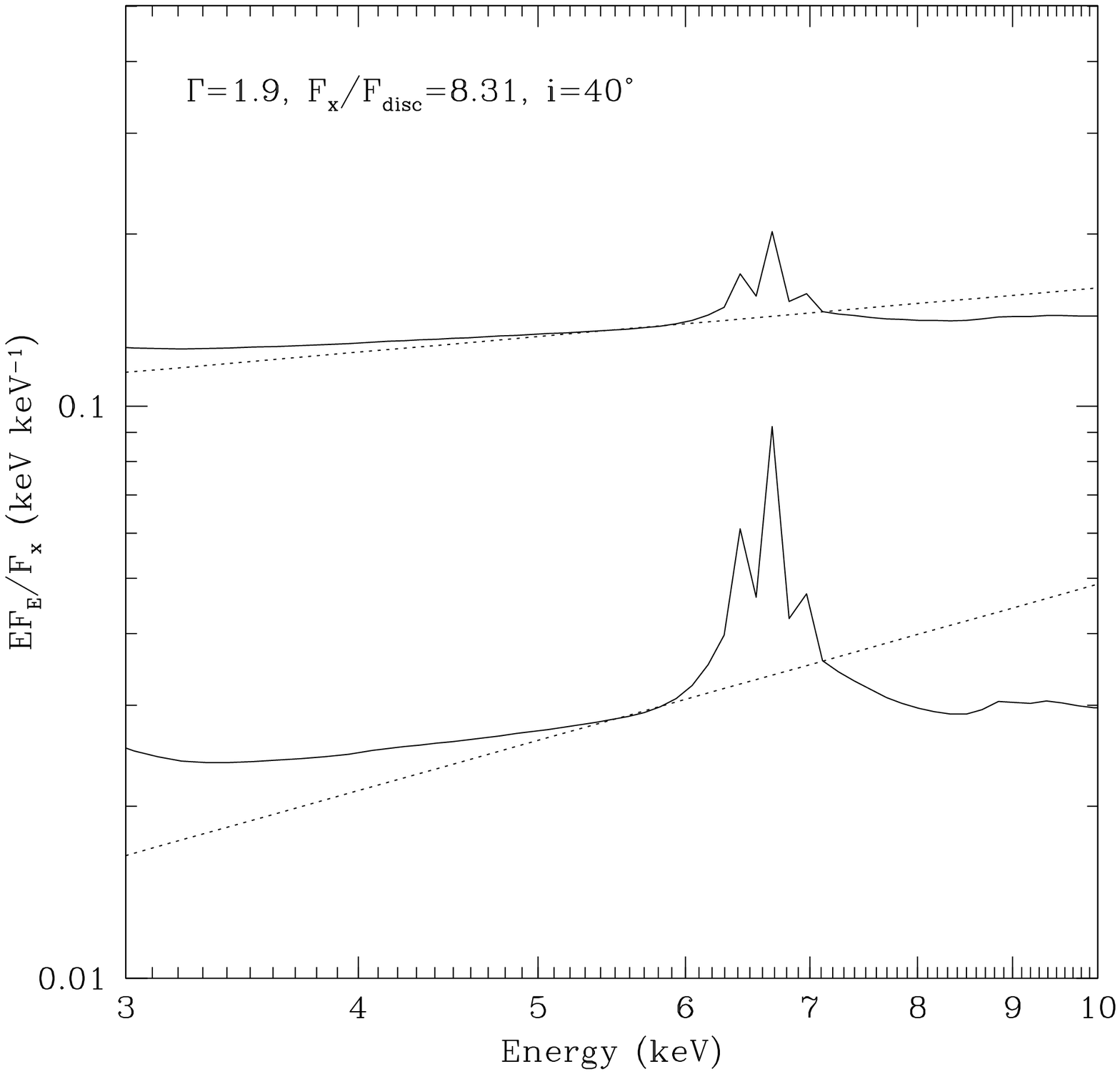}
\caption{An example of the EW and line flux calculation. The lower
  spectrum shows a typical computed reflection spectrum, while the
  upper one has had the illuminating power-law added to it. The
  continuum was estimated fitting a straight line between 5.5 and
  7.1~\kev\ (dotted lines).}
\label{fig:ew-eg}
\end{figure}

The EWs of the \fe\ lines were calculated by performing the following
integral on the total reflected+incident spectrum:
\begin{equation}
\label{eq:ew}
\mathrm{EW} = \int_{5.50~\mathrm{keV}}^{7.11~\mathrm{keV}} { EF_s(E) - EF_c(E)
\over EF_c(E)} dE,
\end{equation}
where $EF_s(E)$ is the spectral flux of the sum, and $EF_c(E)$ is the
estimated spectral flux of the continuum at energy $E$. The continuum
was estimated by fitting a straight line to the reflection spectrum
between 5.50 and 7.11~\kev\ (see Figure~\ref{fig:ew-eg}).  
The line flux was computed by an analogous integral over the same limits.

The lower limit of the integration was chosen to take into account any
Compton broadening of the line profile, and the upper limit was set to
the energy of the photoelectric absorption edge of neutral Fe. Using
the models where \fxfd$=33$, $i=20$~degrees and $\Gamma=$1.5--2.1 a
comparison was made between the EWs calculated by this method and the
ones calculated by \textsc{xspec} using a simple power-law plus
Gaussian model. An \textit{ASCA} SIS-0 response matrix was used to
fake a 500~ks observation for each model spectrum, and the fits were
performed on the simulated 3--10~\kev\ data with the width of the
Gaussian ($\sigma$) fixed at 0.25~\kev. The EWs calculated by
\textsc{xspec} (43--264~eV) were only $\sim 50$\% larger than those
obtained from the above method (26--156~eV). Larger EWs would result
if the reflection fraction was greater than unity.

\begin{figure*}
\begin{minipage}{180mm}
\includegraphics[width=0.5\textwidth]{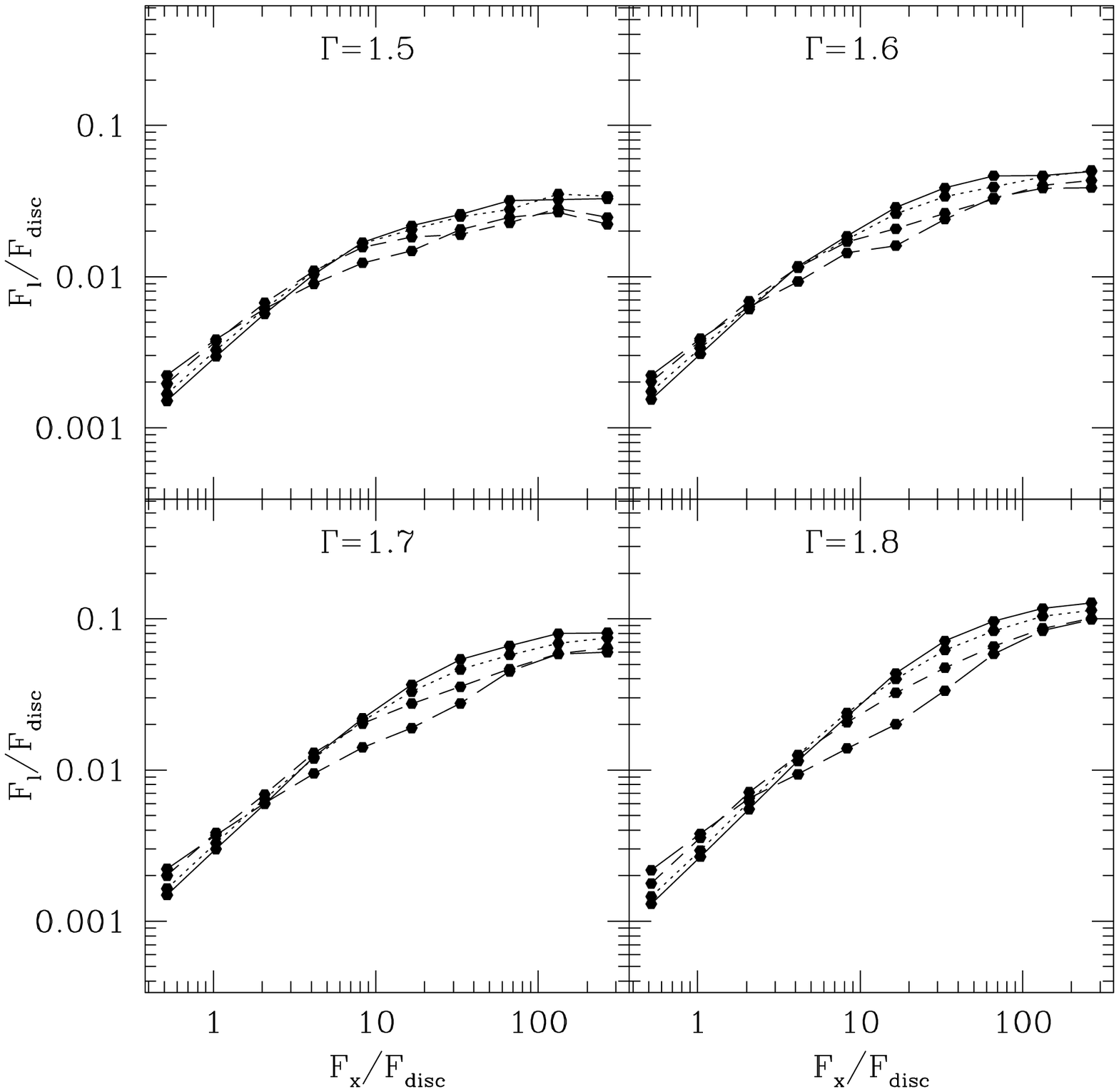}
\includegraphics[width=0.5\textwidth]{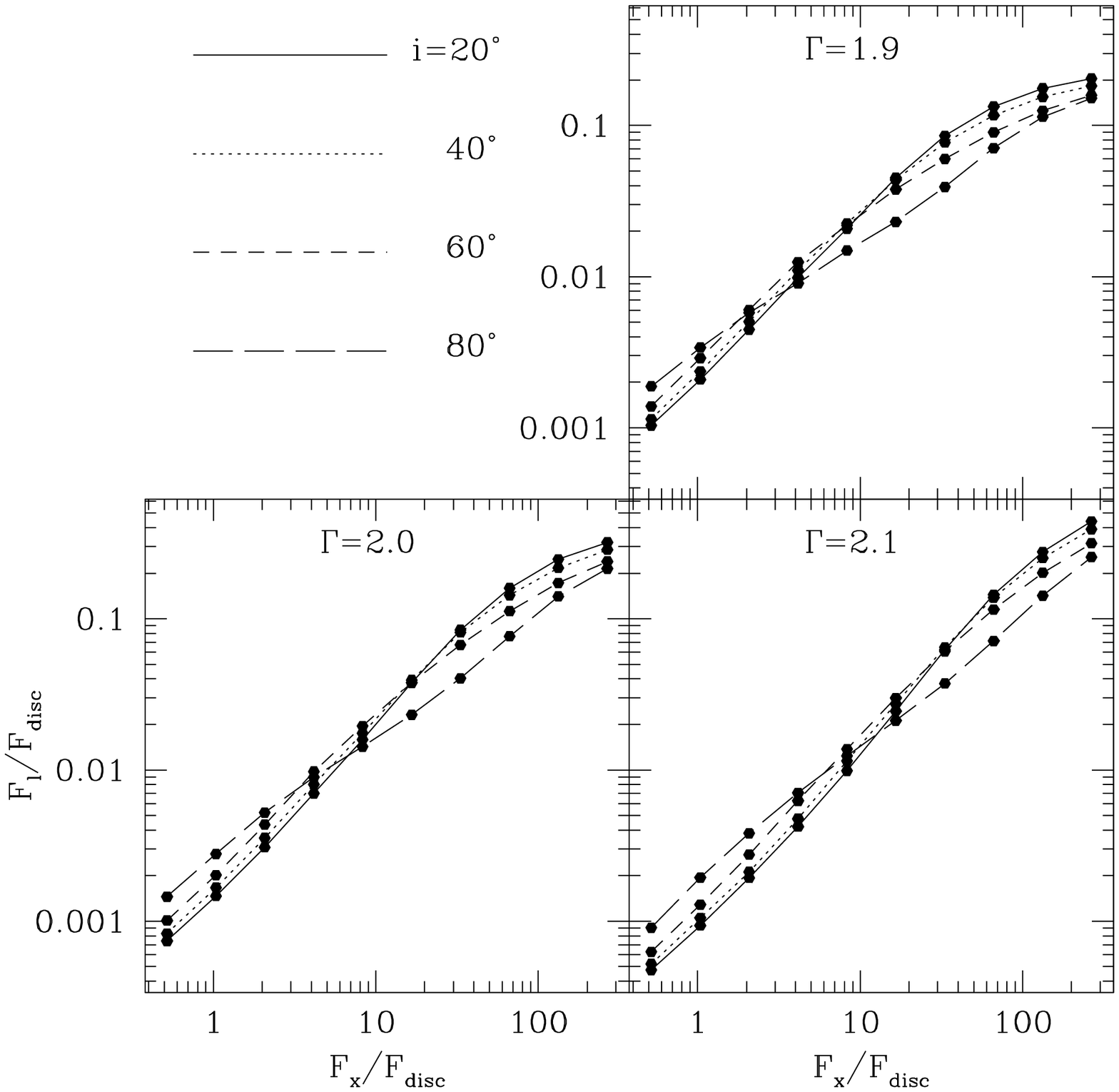}
\caption{Plots of the \fe\ line flux (normalized to
  $F_{\mathrm{disc}}$) versus \fxfd. The separate panels differentiate between the value assumed
  for the photon index, $\Gamma$, of the incident power-law
  spectrum. As indicated, the lines in each panel distinguish between
  different values of the radiation incidence angle. In all cases the
  line flux correlates with the illuminating flux, but the dependence
  is much weaker when the line is highly ionized.}
\label{fig:lineflux}
\end{minipage}
\end{figure*}

Figure~\ref{fig:lineflux} shows how the flux in the \fe\ line depends
on the parameters of the illuminating radiation. Each panel plots the
line flux (normalized to $F_{\mathrm{disc}}$) versus \fxfd\ for the
four different incident angles at a specific photon index.
We find that the line flux is directly proportional to the flux
incident on the disc since the amount of energy emitted by the
atmosphere must follow the amount that it absorbs. However, the
correlation flattens as the line becomes highly ionized and very
weak. This is noticeable for $\Gamma \leq 1.7$. An interesting feature
about this plot is that it illustrates roughly at which \fxfd\ optical
depth effects become important. At low \fxfd\ the line flux increases
with increasing incidence angle because line photons are produced
closer to the surface and can easily escape \citep{geo91}. But this
trend reverses at higher values of \fxfd\ where there is a thicker
ionized skin and ionizing photons are easily scattered out of the
layer before reaching any neutral material. The amount of photons
``lost'' to scattering increases with $i$ and so the line flux drops
as the incidence angle increases. The value of \fxfd\ at which the
crossover occurs increases with $\Gamma$ because greater fluxes are
needed to produce a deep enough ionized layer for the softer incident
spectra.

\begin{figure*}
\begin{minipage}{180mm}
\includegraphics[width=0.5\textwidth]{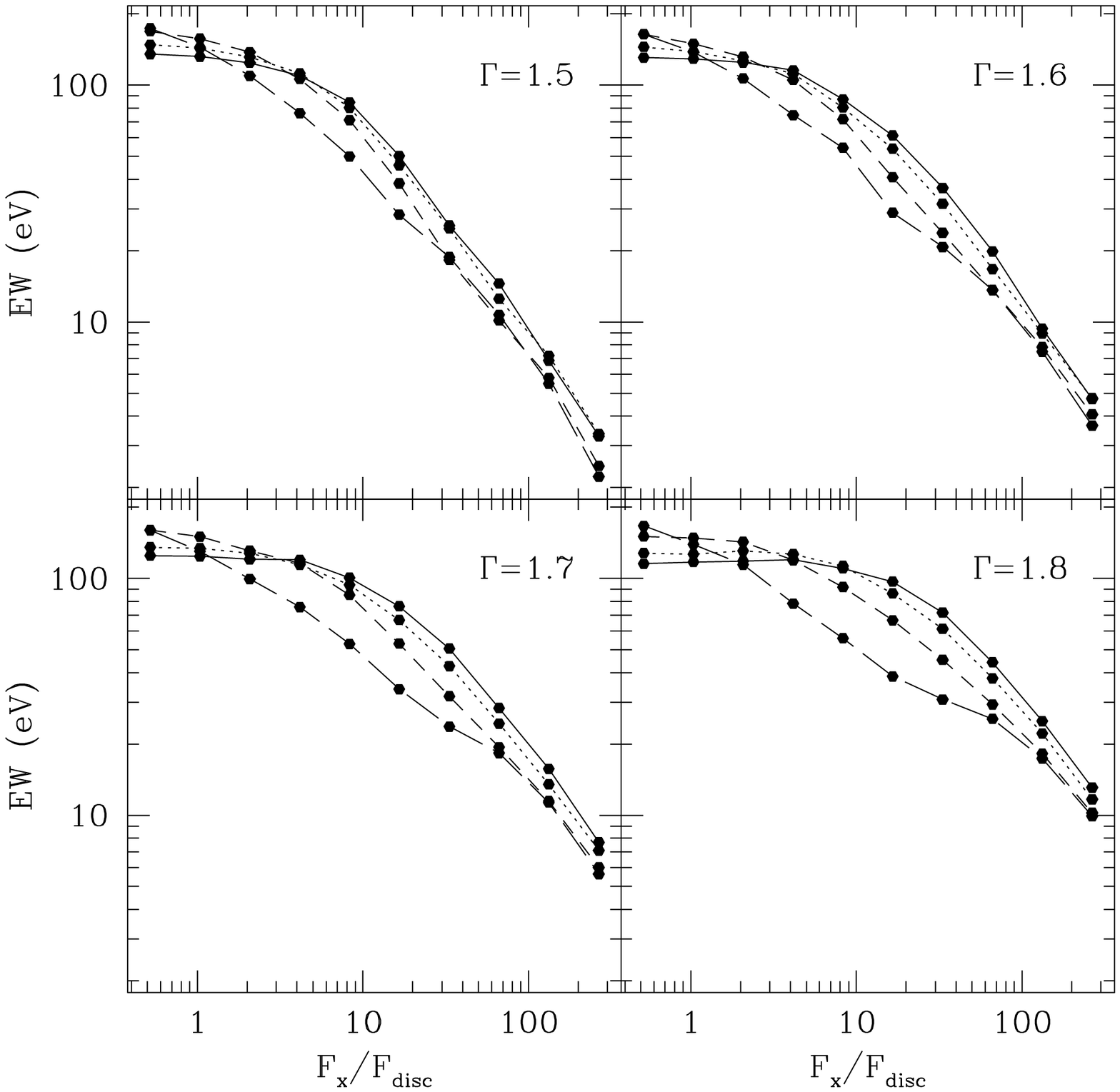}
\includegraphics[width=0.5\textwidth]{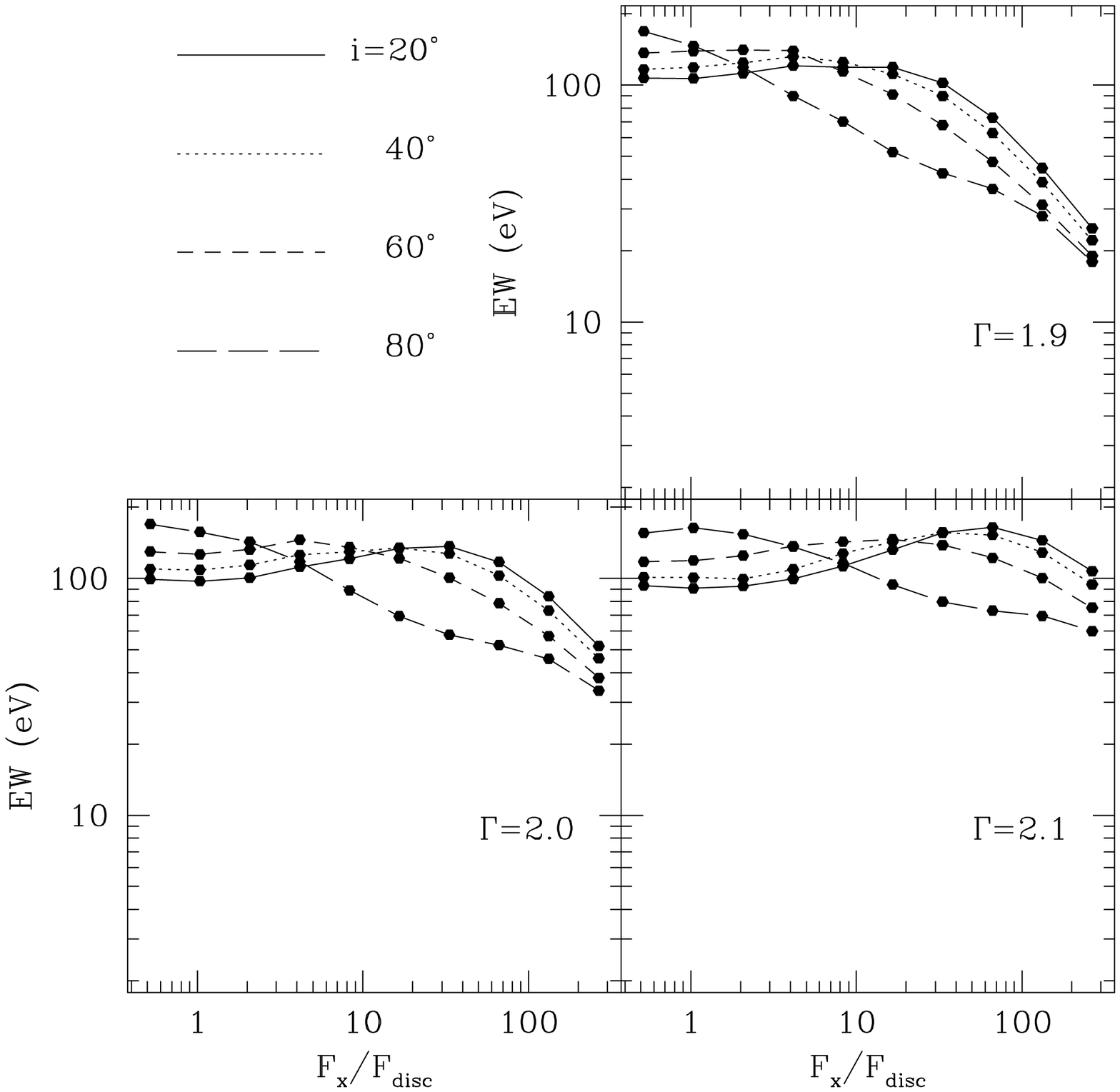}
\caption{Plots of the \fe\ line equivalent width (EW) versus
\fxfd. The separate panels differentiate between the value assumed for
the photon index, $\Gamma$, of the illuminating power-law spectrum. As
indicated, the lines in each panel distinguish between different
values of the radiation incidence angle. For hard irradiating spectra,
the EW will quickly decrease to very small values as the line is
ionized and weakened. The decrease is postponed for softer continua
where a strong He-like line in the reflection spectrum will keep the
EW high.}
\label{fig:ew-flux}
\end{minipage}
\end{figure*}

Turning to the measurements of the equivalent width,
Figure~\ref{fig:ew-flux} plots the \fe\ EW versus \fxfd.
At low \fxfd\ a strong neutral \fe\ line is present in all the
reflection spectra, so the EWs are high. As \fxfd\ is increased the
atmosphere becomes ionized, and where the ionization instability
operates ($\Gamma=1.5$ \& 1.6) no other line replaces the neutral line
so the EW rapidly becomes very small. For softer illuminating spectra,
a He-like line will replace the neutral line, and the EW can remain
high, but, as \fxfd\ grows, this line will ultimately weaken and the
EW will decrease. As seen in Fig.~\ref{fig:lineflux} the
incidence angle crossover marks the onset of an optically thick
ionized layer at the surface of the atmosphere. This point moves to
higher \fxfd\ with larger $\Gamma$. Finally, it is worth emphasizing
that for $\Gamma \geq 1.8$ the \fe\ EW can stay roughly constant over
a factor of about 10 in \fxfd.

\begin{figure}
\includegraphics[width=0.50\textwidth]{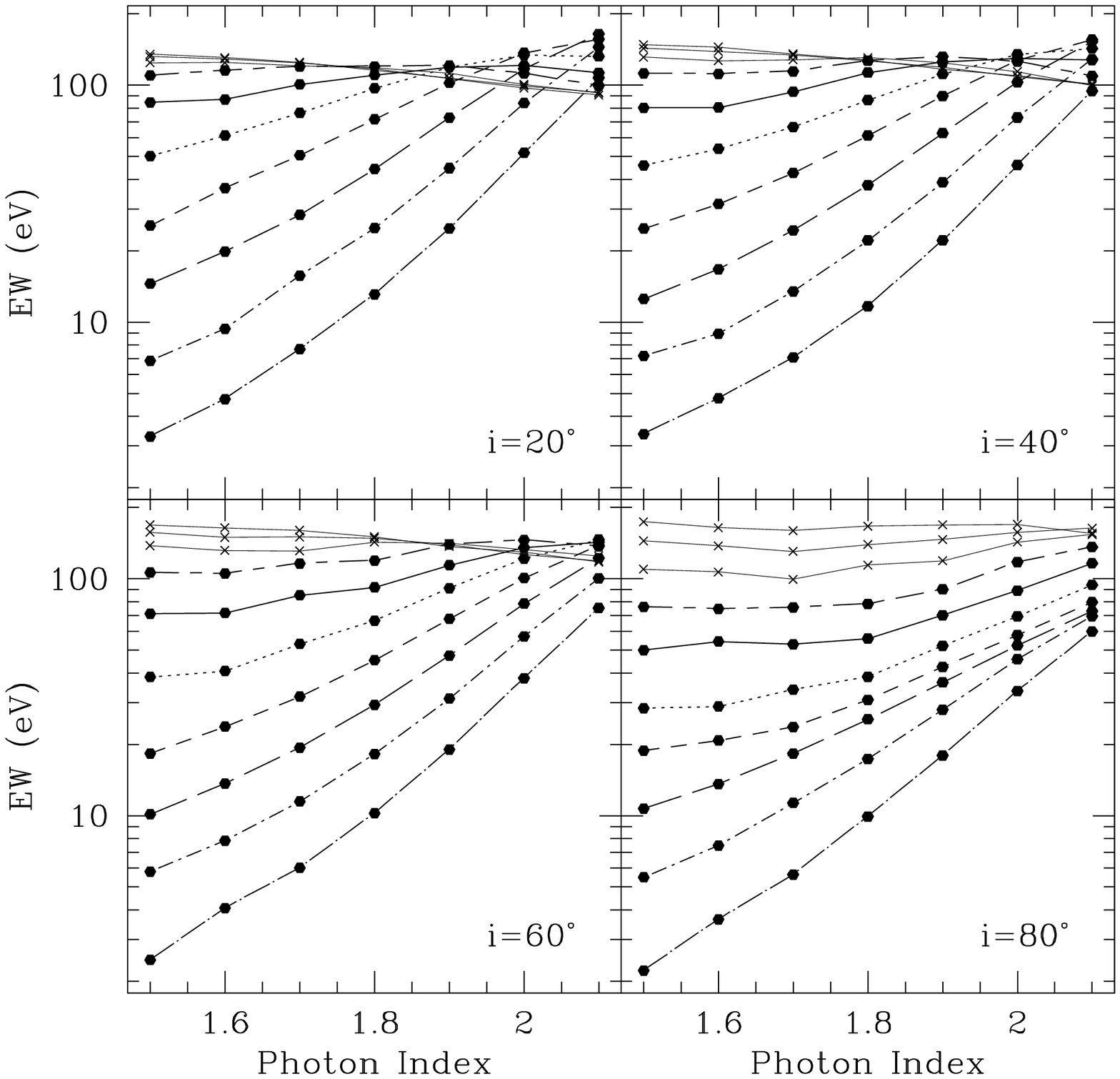}
\caption{The EW of the \fe\ complex plotted against the photon index
of the illuminating power-law. The thin solid lines with the crosses
denote models with \fxfd$ \leq 2.08$, while the other lines with the
solid points (starting with the upper short-dash-long-dash line) trace
out the remaining models with \fxfd\ increasing downwards. When there
is a significant ionized component to the line the correlation is
opposite to the one if the line is predominantly neutral.}
\label{fig:ew-gamma}
\end{figure}

One other interesting correlation that can be shown with this dataset
is illustrated in Figure~\ref{fig:ew-gamma}.
In this plot the models with \fxfd$ \leq 2.08$ are shown with thin
solid lines and crosses. In these cases the \fe\ line complex is
dominated by the neutral line for most values of $\Gamma$, and there
is only a weak dependence of the EW on the photon index. However, when
there is a strong ionized component to the line (the curves with solid
points), there is a significant positive correlation between the \fe\
EW and $\Gamma$. This is a possible explanation for the tentative
EW-$\Gamma$ correlation in the sample of \citet{lub01}.  Furthermore,
it may be possible to use these variations as an independent
observational diagnostic to determine the ionization state of the
accretion disc when other effects (e.g., relativistic broadening)
makes it difficult to estimate the energy centroid of the line.

\subsection{On the self-consistency of the results}
\label{sub:consist}
These computations make the zeroth-order assumption that the changes
in the irradiating flux, $F_x$, will not affect the accretion disc
structure. In reality, changes in $F_x$ are a result of the changes in
disc structure, which will also likely affect
$F_{\mathrm{disc}}$. Here, we consider the impact of our assumption on
the evolution of the \fe\ line.

An increase in $F_x$ would most probably result from an increase in
$f$, the fraction of the viscous energy dissipated in the
corona. However, $f$ is very likely to be a function of the accretion
rate $\dot{m}$. The strength of the dependence of these three
parameters on each other is subject to the unknown details of how
energy is dissipated and transported in accretion discs, but the
dependence of the disc structure on $\dot{m}$ and $f$ is well known
\citep[e.g.,][]{sz94,mfr00}. For the low accretion rate assumed for
these calculations, the disc is gas-pressure dominated and the scale
height of the disc $H \propto \dot{m}^{1/5}(1-f)^{1/10}$. A drop in
$H$ will increase the density of the disc, making it more difficult to
produce ionized lines. However, this will only occur if $F_x$ is
strongly dependent on $\dot{m}$ and/or $f$. It is likely that as $F_x$
is made larger by an increase in $\dot{m}$ the disc moves from a
gas-pressure dominated phase to a radiation pressure dominated one. In
this case $H \propto \dot{m}(1-f)$, but is much larger in amplitude
(by a factor of $\sim 10^3$) than in the gas-pressure dominated
regime. Therefore, the density is much lower and ionized \fe\ lines
are easily produced over a wide range of \fxfd\ \citep*[e.g.,
\fxfd=0.48--2.88,][]{brf2001}.

The disc flux $F_{\mathrm{disc}} \propto \dot{m}(1-f)$, and we have
seen that ionized lines are common when \fxfd$ \ga 2$. Therefore, if
$F_x$ has an equal or greater dependence on $\dot{m}(1-f)$ then
ionized lines should remain common. However, as commented above, there
are very few constraints on the relations between $F_x$, $\dot{m}$,
and $f$. Detailed models of energy transport and dissipation in a
disc-corona system are required to fully test the validity of these
calculations.

\section{Discussion}
\label{sect:discuss}

\subsection{The role of the Compton temperature}
\label{sub:instability}

The results presented in the previous section have shown that the
thermal ionization instability can influence the evolution of the \fe\
line in certain circumstances. This instability has been investigated
previously in the context of broad-line region clouds \citep{kmt81}
and irradiated accretion discs \citep{roz99,nkk00}. Here, we
combine the previous work with the insight gained by our large dataset
to determine, with the minimum of assumptions, the relative importance
of the instability on the \fe\ line.

As discussed by \citet{kmt81} and \citet{gfm83}, the ionization
instability is triggered only when the gas temperature is very high
(as seen in Figs.~\ref{fig:gamma-line} \&~\ref{fig:temp-tau}). In a
photoionized gas in thermal equilibrium the maximum temperature
obtainable\footnote{In reality, the maximum temperature is less than
this limit due to cooling by bremsstrahlung radiation.} is the Compton
temperature $T_\mathrm{C}$:
\begin{equation}
4kT_\mathrm{C} \int^{E_{h}}_{E_{l}} u_E dE = \int^{E_{h}}_{E_{l}} u_E E dE,
\label{eq:ct}
\end{equation}
where $u_E$ is the spectral energy density of the incident radiation
field which extends from $E_{l}$ to $E_{h}$. However, this equation
overestimates $T_\mathrm{C}$ when $E_{h} \ga 30$~\kev\ because it does
not include the Klein-Nishina reduction in Compton heating at higher
energies. For radiation in the form of a single power-law $u_E \propto
E^{1-\Gamma}$, making the first order Klein-Nishina correction
\citep[e.g.,][]{ros99} results in
\begin{equation}
kT_\mathrm{C} = {(2-\Gamma) E_{l} \over 4 (\varepsilon^{2-\Gamma}
-1)} \left [ {\varepsilon^{3-\Gamma} \over (3-\Gamma)} - {21 E_{l}
\varepsilon^{4-\Gamma} \over 5 m_e c^2 (4-\Gamma)} \right ],
\label{eq:ct2}
\end{equation}
where $\varepsilon \equiv E_h/E_l$ and $\Gamma \la 2.5$. This equation
can safely determine $T_{\mathrm{C}}$ for $E_h < 100$~\kev.

The dominant factors in determining $T_{\mathrm{C}}$ are
$\varepsilon^{3-\Gamma}$ and $\varepsilon^{4-\Gamma}$, thus a larger
value of $\Gamma$ results in a lower Compton
temperature. Alternatively, $T_{\mathrm{C}}$ rises if the irradiating
spectrum is cut off at larger values of $E_l$ due to the reduction in inverse
Compton cooling. For example, with $E_h=100$~\kev\ and $\Gamma=2$,
$kT_{\mathrm{C}}=1.3$, 1.6 \& 2.1~\kev\ at $E_l=1$, 10 \& 100~eV. This
result and Fig.~\ref{fig:temp-tau} implies that the lower-limit to the
X-ray power-law would have to cut off in the hard X-rays, contrary to
observations, in order to increase the temperature to high enough
values to initiate the instability when $\Gamma=2$.

Other effects not described by Eq.~\ref{eq:ct2} can also change
$T_{\mathrm{C}}$. As already noted in Sect.~\ref{sub:lines},
$T_{\mathrm{C}}$ rises with an increase in the incidence angle
\citep{nkk00}, although this can only suppress ionized Fe lines when
$\Gamma \sim 2$ if \fxfd$ \ll 1$ where they are unlikely to occur for
most conditions. Also, $T_{\mathrm{C}}$ can decrease
if the disc flux increases due to, say, an increase in accretion rate
\citep{nk01}. The rise in disc flux increases the amount of Compton
cooling and is, in a sense, equivalent to steepening the incident
spectrum.

Observations of a small sample of Seyfert~1 galaxies by
\textit{BeppoSAX} have detected high-energy cutoffs in the power-law
continuum between 70 and 300~\kev\ \citep{m00}. Increasing the cutoff
energy of the incident power-law will increase the Compton temperature
at any value of $\Gamma$ (possibly to the point of initiating the
ionization instability; \citealt{nkaz01}), but it is difficult to
estimate the value of $T_{\mathrm{C}}$ for $E_{h} > 100$~\kev\ as
higher-order Klein-Nishina corrections need to be included. We are
therefore placed in the interesting position where the strength of an
object at 200~\kev\ can affect its spectral features at
6--7~\kev. Models of thermal Comptonization show that the cutoff
energy ($\approx 2kT_e$ of the corona) decreases with larger $\Gamma$,
but the exact form of the relationship depends on the geometry
\citep[e.g.,][]{sve96}. If AGN with $\Gamma$=1.8--2.1 are found to
have spectral cutoffs $< 200$~\kev\ then it is likely that these
objects will exhibit ionized \fe\ lines. However, it is clear that the
calculations need to be extended to higher energies to make more
general predictions, and we intend to do this in future work. It is
also necessary to obtain more measurements of the overall spectral
shape of Seyfert~1 galaxies.

\subsection{Comparison with observations}
\label{sub:compare}

In principle, it is possible to compare the observed long time-scale
variations of the broad \fe\ line in AGN with the model
predictions. There have been many recent studies on \fe\ variability
based on \textit{ASCA} or \textit{RXTE} data \citep[see references
listed in Sect.~\ref{sect:intro} or, for a recent
review,][]{crey01}. However, recent \textit{XMM-Newton} and
\textit{Chandra} results have shown that many AGN have narrow \fe\
components in their spectra \citep{kas01,yaq01,ree01,pou01}. These
lines are expected in AGN unification scenarios where reflection of
X-rays may occur from the inner edge of a molecular torus
\citep*[e.g.,][]{kro94}. It thus seems likely that narrow \fe\ lines
may be common in Seyfert galaxies. These components would have been
unresolvable in \textit{ASCA}, and so the published variations in any
broad line would be contaminated by this feature.

Nevertheless, it may be possible to compare our predictions to the
observations of \mcg. \citet{iwa99} reported that the broad \fe\ line
profile of this Seyfert 1 galaxy shifted dramatically to the red
during a flare event. There was no noticeable narrow component ($EW <
60$~eV) remaining at 6.4~\kev, suggesting that \mcg\ may harbour only a
very weak narrow component. A subsequent long ($\sim 400$~ks) observation of
\mcg\ by \textit{RXTE} reported by \citet{lee00} has resulted in a
comprehensive study of the \fe\ variability. \citet{lee00} split their
data into 4 different continuum flux levels and found that the EW of
the \fe\ line (as fitted by a Gaussian model) dropped as the flux
increased, but that the line flux remained the same. However, if more
complex continuum models that included reflection were used, the line
flux decreases with increasing continuum flux. A re-analysis of this
\textit{RXTE} data by \citet{ve01} found that the line flux varied
with time, but it was not correlated with the continuum. As with
\citet{lee00}, \citet{ve01} found that the photon index was correlated
with the 2--10~\kev\ flux. Finally, \citet{rey00} used sophisticated
numerical techniques on the \textit{RXTE} data to search for a
correlation between the \fe\ line and the continuum over a wider range
of timescales than covered by \citet{lee00}. The search was
unsuccessful which suggests that the line flux is uncorrelated with
the continuum down to timescales of 0.5~ks.

\begin{figure}
\includegraphics[width=0.50\textwidth]{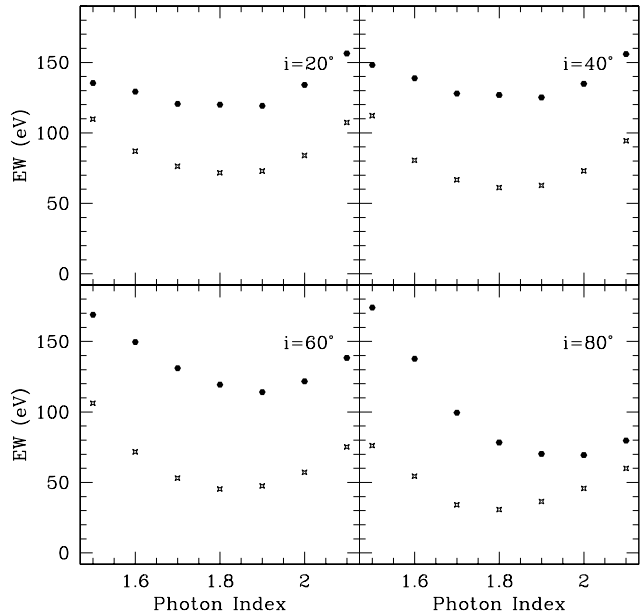}
\caption{The \fe\ EW plotted against the photon index of the
illuminating irradiation, but with \fxfd\ increasing as $\Gamma$
increases. The solid points denote the evolution when \fxfd$=0.52$ at
$\Gamma=1.5$, while the stars show the changes when \fxfd$=4.16$ at
$\Gamma=1.5$. In both cases, \fxfd\ doubles each time $\Gamma$
increases by 0.1.}
\label{fig:ew-photon-fx}
\end{figure}

Figure~\ref{fig:lineflux} predicts that the \fe\ line flux should
correlate with the flux striking the disc. Clearly, this does not
agree with the observations. The EW was observed to drop with the
continuum flux, which does agree with the predictions of
Fig.~\ref{fig:ew-flux}. However, the continuum flux in \mcg\ is
observed to correlate with the photon index, which has the opposite
effect on the EW (Fig.~\ref{fig:ew-gamma}). To give an idea of which
effect might dominate the evolution of the \fe\ EW, we plot in
Figure~\ref{fig:ew-photon-fx} how the EW varies with $\Gamma$, but
tying \fxfd\ to the photon-index.  
The solid points shows the EW variations when \fxfd$=0.52$ at
$\Gamma=1.5$, but with \fxfd\ doubling for an increase in $\Gamma$ of
0.1. In this case the line is mainly neutral and the EW initially
drops due to the increase in $\Gamma$ (recall that increasing \fxfd\
does not change the EW by much when the line is dominated by a
6.4~\kev\ line), before slowly increasing as the ionized component
comes in. The decrease continues the longest when $i=80$ degrees. The
open stars begin at \fxfd$=4.16$ when $\Gamma=1.5$ and trace the
ionized evolution. All four panels show the same decrease-increase
behaviour with the absolute value decreasing with incidence angle. From
these curves and the photon-index range of \mcg\ (1.8--2.1) we can
conclude that the EW evolution of \mcg\ is consistent with that of a
single purely neutral line, or of a hybrid line that is excited by
radiation at a large incidence angle. However, this cannot explain the
strange variation in line flux which may be indicative of more than
one emission component. Indeed, both types of models can fit the data
\citep{bf01}.

\subsection{Neutral and ionized \fe\ lines}
\label{sub:ionized}

The calculations presented here predict that ionized \fe\ lines at
$\sim$ 6.7~\kev\ should be common in almost all cases if \fxfd$ \gg
1$. Evidence for ionized lines have been seen in the X-ray spectra of
some Narrow Line Seyfert 1 Galaxies
\citep*{com98,com01,tgn98,vau99,bif01,tjy01,trg01} and high-luminosity
Seyferts/low-luminosity quasars \citep*{rov00,pou01,orr01}, both of
which are thought to be high accretion rate objects and so might not
be directly applicable to the current computations. Most Seyfert
galaxies seem to exhibit lines of neutral Fe at 6.4~\kev\
\citep[e.g.][]{n97a}, although this may be a result of contamination
from a narrow component. Taking these results at face value suggests
that discs are either weakly illuminated (i.e., \fxfd$< 1$) or much
denser than previously considered.

If discs are weakly illuminated then it is difficult to reconcile this
with the otherwise very successful theory of coronal heating by
magnetic flares which necessarily predicts a highly irradiated
disc. It is possible that measurements of the centroid energy of the
\fe\ line can be confused by poor signal-to-noise data, calibration
problems, or gravitational redshifting effects. Searching for EW
variations with the photon index through relations shown in
Fig.~\ref{fig:ew-gamma} may provide an alternative test for
determining the ionization state of the \fe\ line.

Extensions to the basic magnetic flare model can also reconcile the
models with the data. For example, \citet{nkk00}, \citet{nay00} and
\citet{dn01} discuss the idea that a local wind will be driven by a
magnetic flare event. These authors argue that as a result of this
wind, surface material on the disc will be blown away revealing denser
gas deeper into the disc. It is this gas that produces the observed
reflection spectrum complete with a neutral \fe\ line. However, the
effects (or even the existence) of such an X-ray driven wind has not
been properly modelled, so, in light of the results of this paper, the
following model is proposed as an alternative.

\begin{figure}
\includegraphics[width=0.5\textwidth]{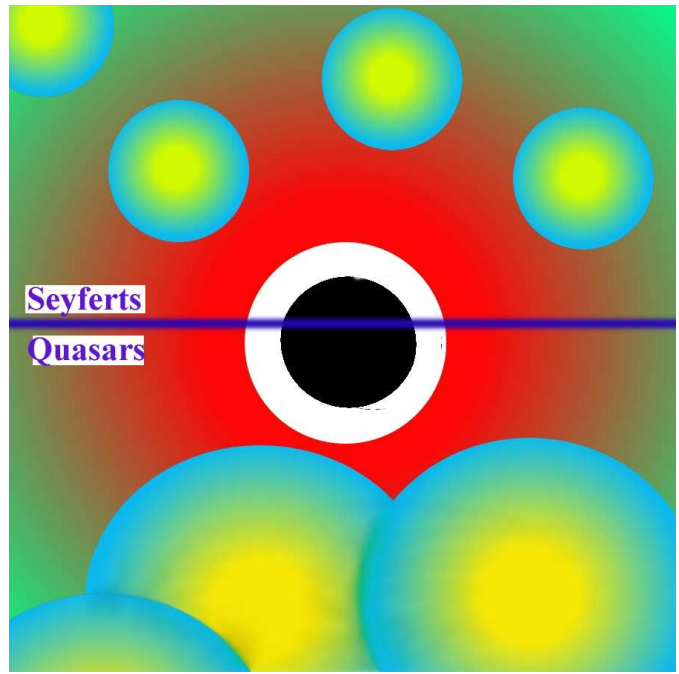}
\caption{A schematic picture of how ionization gradients may effect
  observations of the \fe\ line. This illustration shows from a polar
  viewpoint an accretion disc (large red to green gradient) around a
  black hole. On top of the accretion disc magnetic flares are
  occurring (yellow-blue regions). In the top half of the figure the
  illumination events are fairly small and localized. Therefore,
  although they may highly ionize the gas directly below them (yellow
  regions), the amount of area that is producing strong ionized iron
  lines at 6.7~\kev\ (blue regions) is relatively small compared to
  the area of the disc that is weakly illuminated. Therefore, a strong
  neutral line will be observed. This description might be
  appropriate for lower-luminosity objects such as Seyferts. The
  bottom half of the figure shows more powerful and (perhaps) more
  numerous magnetic flares. In this case, there are large areas of
  the disc that are completely ionized and so only a weak \fe\ line
  may be detected. This could explain the X-ray Baldwin effect in
  quasars if the higher luminosity in these objects translates into a
  more active corona.}
\label{fig:schematic}
\end{figure}

Consider the situation when separate magnetic flares are occurring
over the accretion disc \citep[see
Figure~\ref{fig:schematic} and][]{mf01}.
Directly underneath each illuminating event, the surface of the disc
is highly ionized and therefore produces no \fe\ emission. As the
angle of incidence of the radiation increases, the flux incident on
the gas drops resulting in ionized \fe\ lines. As the incidence angle
further increases, the local \fxfd\ decreases until eventually neutral
lines dominate. Such a gradient in ionization on the disc has also
been considered by \citet{rey00} for AGN and by \citet{dz99},
\citet{yfr01} and \citet{dn01} to explain the low reflection fractions
observed in low-state Galactic Black Hole Candidates. Since there is
much more surface area on the disc that is weakly illuminated and/or
illuminated at grazing incidence, the majority of the line emission
detected by a telescope will be at neutral energies. This effect
should work for any value of $\Gamma$, as the combination of low
\fxfd\ and large $i$ will result in a neutral line
(Fig.~\ref{fig:angle-line}). Similarly, the complete ionization below
each flare can occur at any photon-index, but its assurance will
depend on the physical and geometrical details of the event itself.

Spectral evolution of the illumination events will also cause changes
in the ionization state of the \fe\ line. The recent \textit{RXTE}
observation of the black hole candidate GX~339-4 reported by
\citet{fen01} showed that as the X-ray flux decreased from its low
state to an ``off state'', its \fe\ line evolved from neutral at
6.4~\kev\ to ionized at 6.7~\kev. More importantly, the photon index of the
continuum evolved from about 1.5 to 2.2 as the flux decreased. As
discussed earlier (Sect~\ref{sub:instability}), when $\Gamma=1.5$ the
thermal ionization instability can operate and a neutral line is
expected, but it is highly unlikely to work when $\Gamma=2.2$, and a
6.7~\kev\ line would be observed. Perhaps this is an observation of a
decaying and expanding flare.

\subsubsection{X-ray Baldwin effect}
\label{subsub:baldwin}
Observations of many Seyfert galaxies and quasars by \textit{ASCA}
have shown that as the X-ray luminosity of the source increases, the
EW of the \fe\ line decreases until it becomes undetectable at a
2--10~\kev\ luminosity of 10$^{46}$~erg~s$^{-1}$
\citep{iwa93,n97b,ree00}. This trend has been named the X-ray Baldwin
effect in analogy to the decrease in EW of the C~{\sc
iv}~$\lambda$1550 line in the UV continuum of QSOs \citep{bal77}. In
terms of the constant density models, this effect could easily be
explained as the ionization parameter of the gas increasing to the
point where the outer layers of the disc are fully ionized and
therefore would not imprint any line emission on the reflection
spectrum. The ionization parameter would increase if the higher
luminosity in quasars was due to a higher accretion rate
\citep{mfr93}. Newer hydrostatic disc models also can explain the
X-ray Baldwin effect by ionizing a thick skin on the surface of the
accretion disk \citep{nay00,zyk01}.

As discussed earlier, ionization gradients along the accretion disc by
different flare locations may be able to account for the observed
energies of the \fe\ line. If the power and number of flares above the
disk increases as one moves from Seyferts to quasars then this would
increase the X-ray luminosity, and also increase the amount of surface
area that is subject to a highly ionizing flux (see
Fig.~\ref{fig:schematic}). In such a situation only a very weak,
highly ionized \fe\ line would be emitted from the surface. Indeed,
recent \textit{XMM-Newton} observations show that ionized lines are
observed in higher-luminosity objects \citep{ree01,pou01}.

Such a scenario is not too different from that envisaged by the other
models, but suggests a more complex geometry. There are a number of
details to work out in order for such a model to placed on firm
theoretical ground. Clearly, in a situation such as the one presented in
the lower part of Fig.~\ref{fig:schematic}, feedback effects between
the reflection spectrum and the coronal activity would be vitally
important and may even assist in erasing spectral features through
Comptonization \citep{pet01}. 

\section{Conclusions}
\label{sect:concl}
The broad \fe\ line which (most likely) results from reflection from
an accretion disc is the most important spectral signature in the
X-ray study of AGN. The model calculations presented in this paper
have shown that the behaviour and properties of the Fe line from an
irradiated disc in equilibrium depend on how it is
illuminated. Specifically, when the incident flux is larger than the
disc flux, we predict that a He-like \fe\ line at 6.7~\kev\ will be a
common feature in the reflection spectrum. The thermal ionization
instability is able to change this and produce a neutral line at 6.4~\kev,
but only when $\Gamma \leq 1.6$ (for a cutoff energy of 100~\kev) and
the incident flux is not too high.

The \fe\ EW will remain roughly constant as \fxfd\ is increased until
the atmosphere becomes highly ionized and then it will decrease to
small values. The rate of decline is lower for higher values of
$\Gamma$. When the line complex is dominated by He-like Fe, the EW
will strongly increase with $\Gamma$, but will slightly decline with
the photon-index if it is neutral. This effect could be used as an
observational diagnostic to determine the ionization state of the
line, independent of the line centroid.

Long observations of Seyferts \& quasars by \textit{XMM-Newton} would
be required to test the predictions of the models. However, any narrow
component to the line profile must be taken into account before a
comparison could be made. The results of such a study would be
important, as it is clear that the ionization state of the disc
depends greatly on how it is irradiated which would greatly constrain
coronal models.

The prediction that ionized lines are common can be reconciled with
their observational rarity by employing multiple ionization gradients
on the accretion disc, which would be a natural result of a
magnetically active and patchy corona. The challenge is now extended
to modellers to compute observationally testable predictions for such
complex scenarios.

\section*{Acknowledgments}
The authors thank A.\ Fabian, S.\ Vaughan, K.\ Iwasawa \& A.\ Merloni for
valuable discussions, and C.\ Berger for assistance with
the models.  DRB acknowledges financial support from the
Commonwealth Scholarship and Fellowship Plan and the Natural Sciences
and Engineering Research Council of Canada. RRR acknowledges support
from the College of the Holy Cross.


\bsp 

\label{lastpage}

\end{document}